\documentclass[a4paper,fleqn]{cas-sc}

\usepackage[numbers]{natbib}
\usepackage{amsmath,amsfonts,amssymb}
\usepackage{array}
\usepackage{booktabs}
\usepackage{multirow}
\usepackage{tabularx}
\usepackage{ragged2e}
\usepackage{url}
\usepackage{graphicx}
\usepackage{float}
\usepackage{longtable}
\usepackage[section]{placeins}

\begin{document}
\let\WriteBookmarks\relax
\def\floatpagepagefraction{1}
\def\textpagefraction{.001}

\shorttitle{Review on Electric Railway System Optimization}
\shortauthors{Liu et al.}

\title[mode=title]{Review on Electric Railway System Optimization: Train Dynamic Scheduling, Energy Management, and Storage Integration}

\tnotemark[1]
\tnotetext[1]{This work was supported by Europe's Rail Flagship Project 4 -- Sustainable and green rail systems, and Trafikverket.}

\author[1]{Fei Liu}
\credit{Conceptualization, Writing -- original draft}

\author[2]{Can Wan}
\credit{Writing -- review \& editing, Supervision}

\author[1]{Stefan \"{O}stlund}
\credit{Writing -- review \& editing}

\author[1]{Qianwen Xu}
\credit{Writing -- review \& editing, Supervision}

\affiliation[1]{organization={KTH Royal Institute of Technology},
            city={Stockholm},
            country={Sweden}}

\affiliation[2]{organization={Zhejiang University},
            city={Hangzhou},
            country={China}}

\begin{abstract}
The modernization of railway systems is being driven by the need for greater efficiency, sustainability, and intelligent operation. In this review, recent advancements in dynamic scheduling, energy management, and energy storage systems (ESSs) integration within electric railway networks are analyzed. Optimization strategies for train scheduling and operation control are reviewed, with a focus on methods that reduce energy consumption and improve overall system performance. The role of energy storage technologies is analyzed in terms of energy management, peak shaving, and voltage and frequency control for practical application. The integration of artificial intelligence and advanced control strategies is also reviewed in electrified railway systems. This review provides a comprehensive overview of current research trends and outlines future directions for the development of resilient, energy efficient, and intelligent railway systems with ESSs integration.

\noindent\textit{Word count: $\sim$7,500}
\end{abstract}

%%\begin{highlights}
%%\item Dynamic scheduling, trajectory control, and ESS management are reviewed within a unified framework.
%%\item The interdependencies between timetable decisions, speed profiles, and ESS operation are explicitly identified.
%%\item ESS coupling depth is characterized across joint, coordinated, and decoupled modes, spanning scheduling, control, and application scenarios.
%%\end{highlights}

\begin{keywords}
Railway system \sep Dynamic scheduling \sep Operation control \sep Energy management \sep Energy storage system
\end{keywords}

\maketitle

%%=============================================
\section{Introduction}
\label{sec:intro}
%%=============================================

Railway systems represent a critical component of modern transportation infrastructure, playing a significant role in improving mobility, and promoting environmental sustainability \cite{c1}. As global efforts to improve energy efficiency and reduce carbon emissions intensify, railway networks are undergoing rapid technological advancements aimed at optimizing their design, operation, and management. The European Union has placed railway transformation at the center of its research and innovation agenda through Horizon Europe \cite{c2}, its primary funding program for research and development for the period 2021--2027. A key initiative within this framework is Europe's Joint Rail Undertaking (EU-Rail) \cite{c3}, which aims to drive digitalization and sustainability in the railway sector.

One of the primary challenges in railway modernization is achieving operational efficiency while meeting increasing passenger and freight demands. To manage daily operations efficiently, modern railway operations rely on real-time monitoring and automated control \cite{c4} to optimize traffic flow and adjust schedules dynamically. Centralized management systems integrate train movement tracking, remote route setting, and traffic rescheduling to ensure smooth operations, especially in response to disruptions. The seamless coordination of these technologies enables efficient dynamic scheduling and operational control \cite{c5}, ensuring railways can meet growing demands while maintaining high performance and reliability.

Dynamic traffic scheduling and operation control are two fundamental aspects of modern railway systems, working together to ensure efficiency, safety, and energy optimization. Dynamic scheduling \cite{c6,c7,c8} adjusts train timetables in response to delays, minimizing disruptions while optimizing travel time and energy consumption. Traditional scheduling methods are now enhanced with AI-driven models and optimization algorithms, enabling real-time decision-making that improves operational reliability. Timetable optimization \cite{c9} serves as a bridge between scheduling and train operation, helping to reduce delays while improving energy efficiency. Advanced approaches, such as cooperative train control and speed profile optimization for automatic train operation (ATO), allow trains to adjust running and dwell times dynamically, maximizing the use of regenerative braking energy (RBE) and reducing energy consumption. Beyond scheduling, operation control \cite{c10,c11} ensures that trains maintain safe and efficient trajectories while optimizing energy use and network capacity. Modern control systems rely on real-time train-to-ground and train-to-train communication, enabling automated coordination of multiple trains. The development of advanced computing and control technologies has led to the widespread adoption of ATO \cite{c12}, which automates acceleration, coasting, and braking decisions \cite{c13}. Seamlessly integrating dynamic scheduling, optimized timetables, and real-time operation control is essential for managing the increasing complexity of railway systems. The growing deployment of energy storage systems (ESSs) further extends this integration requirement, given that ESS charging and discharging decisions are tightly coupled with train timetables and speed trajectories.

\begin{figure*}[htbp]
\centering
\includegraphics[width=0.82\textwidth]{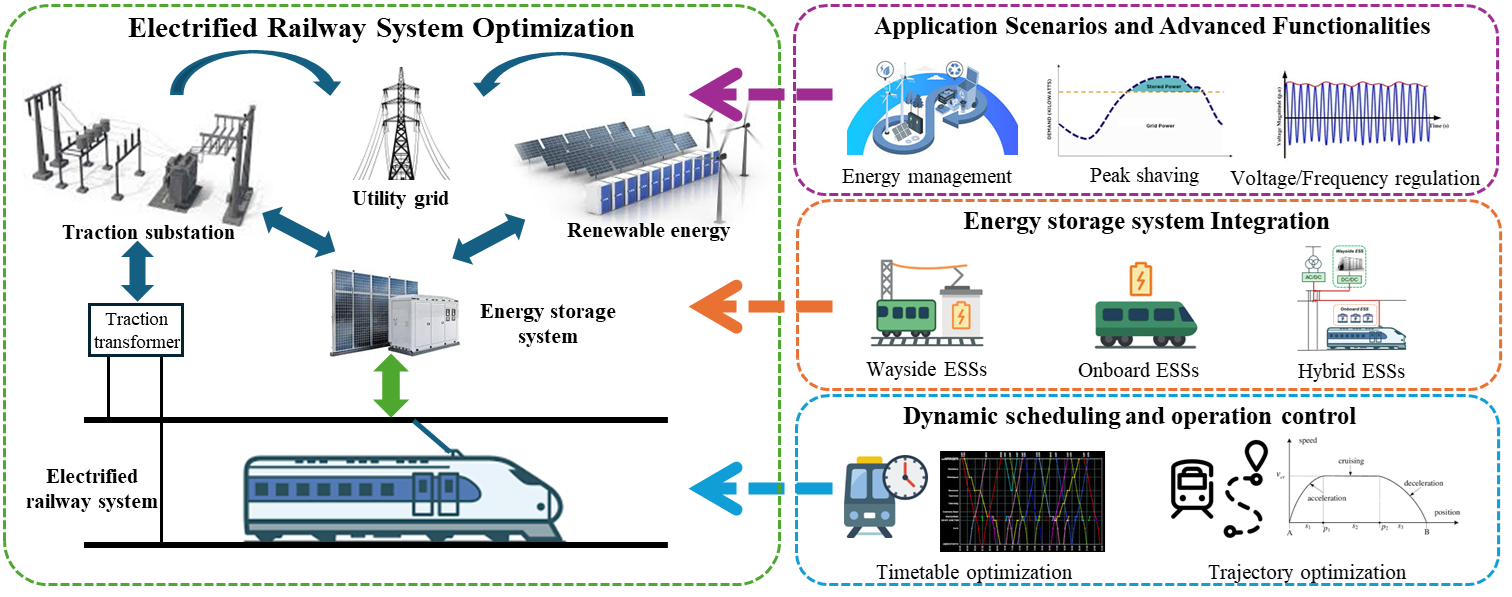}
\caption{Optimization Strategies and System Architecture in Electrified Railway Networks.}
\label{fig:optimization}
\end{figure*}

In addition to improving operational efficiency, modern electric railway systems face growing pressure on their electrical infrastructure due to increasing traffic demand and higher peak power requirements. Electrified railways often generate substantial regenerative energy during braking \cite{c14,c15}, but without appropriate storage solutions, this energy is frequently wasted because of grid limitations or mismatched supply and demand. The effective integration of energy storage systems (ESSs) into railway networks provides a promising pathway to address these challenges. By integrating efficient ESSs such as onboard and wayside ESSs, railway networks can significantly reduce energy losses, enhance grid stability, mitigate peak power, and improve overall system resilience \cite{c16,c17}. Onboard ESS stores energy \cite{c18,c19} directly within the train, providing immediate power for acceleration and smoothing energy fluctuations. Meanwhile, wayside ESS collects excess energy from multiple trains and redistributes it as needed, supporting grid stability and reducing peak power demand \cite{c20}.
To fully realize these benefits, further attention must be given to the control, sizing, and placement of ESSs within railway systems. Intelligent control strategies dynamically adjust charge and discharge cycles in response to real-time power demand, train schedules, and grid conditions \cite{c21}, while careful planning of system capacity and location ensures a balance between performance and cost.

ESS integration in railway networks enables a range of advanced functionalities. In energy management, ESS helps optimize overall power usage by capturing regenerated energy from external electricity sources. It also plays a key role in peak shaving, flattening demand spikes and distributing loads more evenly over time, thereby relieving stress on the power supply system and reducing the risk of voltage fluctuations\cite{c22}. Beyond direct energy savings, ESS improves infrastructure stability and grid interaction. Smart storage systems contribute to voltage \cite{c23,c24} and frequency regulation \cite{c25,c26}, supporting broader grid stabilization efforts \cite{c27}, particularly in electrified networks.

The optimization of electric railway systems has attracted growing research attention, resulting in several dedicated reviews. Online rescheduling and traffic management have been reviewed from an operational reliability perspective~\cite{c_add2}, while energy efficient train control and timetabling have been examined in~\cite{c5,c_add3,c_add4}. Regarding energy and storage technologies, existing reviews have addressed RBE recuperation~\cite{c_add7}, onboard storage systems~\cite{c18,c19}, railway ESS technologies and applications~\cite{c_add5, c_add8,c_add11}, and power flow control based RBE utilization in AC electrified networks~\cite{c_add6}. In addition, artificial intelligence (AI) applications in railway systems have been reviewed from systematic literature and taxonomy-oriented perspectives~\cite{c_add9,c_add10}. 
These works have each advanced understanding within their respective domains. However, the relationships between timetable level decisions, train speed trajectories, and ESS operation have received limited attention as an integrated subject. Moreover, how deeply ESS couples with the scheduling and control layers differs substantially across architectures and application scenarios, a distinction that existing surveys have not systematically drawn. Table~\ref{tab:review_comparison} summarizes the scope of representative reviews relative to the present work.

\setlength{\tabcolsep}{3pt}
\begin{longtable}{|>{\small\sffamily}p{1.8cm}|>{\small\sffamily}p{2.8cm}|>{\small\sffamily}p{2.6cm}|>{\small\sffamily}p{2.2cm}|>{\small\sffamily}p{2.6cm}|>{\small\sffamily}p{2.8cm}|}
\caption{Scope Comparison of Representative Existing Reviews}
\label{tab:review_comparison}\\
\hline
\textbf{Reference} & \textbf{Dynamic Scheduling} & \textbf{Trajectory Optimization} & \textbf{ESS Integration} & \textbf{AI-driven Methods} & \textbf{Cross-layer Coupling Analysis} \\
\hline
\endfirsthead
\multicolumn{6}{l}{\small\itshape (Table~\ref{tab:review_comparison} continued)}\\
\hline
\textbf{Reference} & \textbf{Dynamic Scheduling} & \textbf{Trajectory Optimization} & \textbf{ESS Integration} & \textbf{AI-driven Methods} & \textbf{Cross-layer Coupling Analysis} \\
\hline
\endhead
\hline
\endfoot
\endlastfoot

\cite{c_add2} 
& \checkmark 
& $\times$ 
& $\times$ 
& $\times$ 
& $\times$ \\ \hline

\cite{c_add3} 
& $\times$ 
& \checkmark 
& $\times$ 
& $\times$ 
& $\times$ \\ \hline

\cite{c_add4} 
& Partial 
& \checkmark 
& $\times$ 
& Partial 
& $\times$ \\ \hline

\cite{c5} 
& \checkmark 
& \checkmark 
& $\times$ 
& \checkmark 
& Partial \\ \hline

\cite{c_add7} 
& Partial 
& $\times$ 
& \checkmark~
& $\times$ 
& Partial \\ \hline

\cite{c18,c19} 
& $\times$ 
& $\times$ 
& \checkmark~(onboard) 
& $\times$ 
& $\times$ \\ \hline

\cite{c_add5,c_add8,c_add11} 
& $\times$ 
& $\times$ 
& \checkmark 
& $\times$ 
& $\times$ \\ \hline

\cite{c_add6} 
& $\times$ 
& Partial 
& \checkmark~
& $\times$ 
& Partial \\ \hline

\cite{c_add9,c_add10} 
& Partial 
& Partial 
& $\times$ 
& \checkmark 
& $\times$ \\ \hline

This work 
& \checkmark 
& \checkmark 
& \checkmark 
& \checkmark 
& \checkmark \\ \hline

\end{longtable}
\setlength{\tabcolsep}{6pt}

Motivated by the gaps identified above, this review systematically analyzes recent advances in dynamic scheduling, operation control, and ESS integration for electric railway systems, with an explicit focus on the cross layer interactions that existing surveys have not addressed. The rest of this review is organized as follows. Section~\ref{sec:tpss} introduces the structure of train power supply systems. An overview of dynamic scheduling strategies in railway systems, including optimization algorithms and AI-driven methods for timetable adjustments and delay minimization, is presented in Section~\ref{sec:scheduling}. The integration of ESSs is discussed in Section~\ref{sec:ess}. Application scenarios of energy storage and control strategies, including peak shaving, load balancing, and energy management are explored in Section~\ref{sec:application}. Finally, challenges and future research are addressed in Section~\ref{sec:conclusion}.

%%=============================================
\section{Train Power Supply System Structure}
\label{sec:tpss}
%%=============================================

The train power supply system (TPSS) in electrified railway networks comprises multiple interconnected components responsible for delivering electrical energy from the utility grid to the train. As illustrated in Fig.~\ref{fig:optimization}, the system begins with traction substations, which draw power from the utility grid. These substations include traction transformers that step down high-voltage AC power to suitable levels for train operation. Power is then fed into the electrified railway system via overhead contact lines. The TPSS architecture also incorporates ESSs. These units enhance energy efficiency by recovering regenerative braking energy and supporting peak load reduction. Building on this structural foundation, this review further investigates dynamic train scheduling, energy-efficient operation, and ESS control strategies across energy management, peak shaving, and voltage/frequency regulation scenarios.

%%=============================================
\section{Dynamic Scheduling and Operation Control in Railway Systems}
\label{sec:scheduling}
%%=============================================

Railway systems rely on dynamic scheduling and operation control, two interdependent strategies that enable real-time decision-making and adaptability to network conditions. Dynamic scheduling focuses on adjusting train timetables in response to delays, disruptions, or fluctuations in demand, ensuring efficient resource allocation, train coordination, and passenger flow management. On the other hand, operation control involves regulating train trajectories and optimizing energy consumption. Timetable decisions determine when and how much regenerative braking energy is produced across the network, while speed profiles govern the rate and pattern of traction power demand, establishing the energy input conditions that ESS management must respond to.

\begin{figure*}[htbp]
\centering
\includegraphics[width=0.82\textwidth]{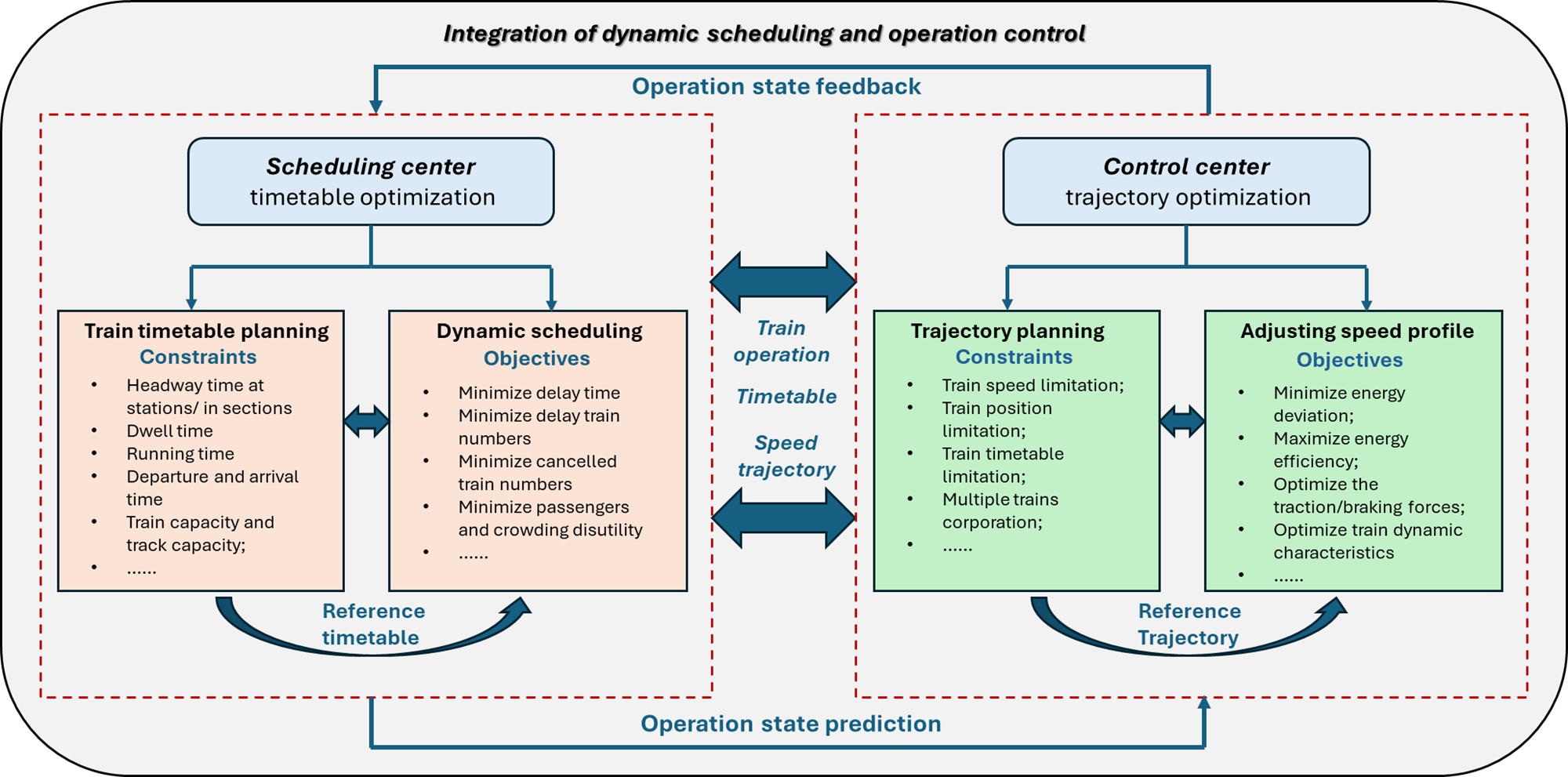}
\caption{The framework of dynamic scheduling and operation control.}
\label{fig:framework}
\end{figure*}

The following sections will explore the methodologies and technologies, detailing the optimization techniques, real-time decision models, and integration frameworks. The whole framework of this section is shown in Fig.~\ref{fig:framework}.

\subsection{Adaptive Train Scheduling and Timetable Optimization}

Maintaining an efficient and reliable railway network requires continuous adjustments of train schedules to accommodate disruptions \cite{c28}, fluctuating demand, and operational constraints. In real-world operations, deviations from the planned timetable are inevitable due to unexpected disturbances, infrastructure limitations \cite{c28}, and network congestion \cite{c29}. To address these challenges, railway systems employ adaptive scheduling strategies that dynamically modify train departure and arrival times, dwell durations, route assignments, and sequencing.

Dynamic adjustments to the train timetable must balance multiple objectives, including reducing overall delay times, minimizing deviations from planned schedules, and maintaining system-wide efficiency \cite{c29}-\cite{c33}. In addition, scheduling models integrate various constraints, such as track capacity, signal block limitations, and operational headways, ensuring that modifications comply with safety and infrastructure requirements. Beyond operational constraints, energy-related considerations such as peak power demand and RBE recovery are increasingly incorporated as additional optimization criteria, reflecting the growing interdependence between timetable design and power system management \cite{c31}.

\subsubsection{Objective functions}

The optimization of train timetables generally aims to balance multiple objectives, which can be classified into the following categories:

Minimizing total delay time: one of the most common objectives in timetable optimization is reducing the cumulative delay of all trains, ensuring that trains adhere as closely as possible to the planned schedule \cite{c29}.
\begin{equation}
\min \sum_{t \in \mathcal{T}} \Delta_t
\end{equation}
where $\mathcal{T}$ is the set of all trains, and $\Delta_t$ denotes the delay of train $t$ relative to its scheduled arrival time.

Minimizing the number of delayed trains: rather than focusing solely on the duration of delays, some models prioritize minimizing the number of affected trains to maintain overall network stability \cite{c29}.
\begin{equation}
\min \sum_{t \in \mathcal{T}} \delta_t
\end{equation}
where $\delta_t$ is a binary variable equal to 1 if train $t$ is delayed beyond a given threshold, and 0 otherwise.

Minimizing passenger travel time: passenger travel time models aim to reduce total journey duration, including waiting times, transfer times, and onboard travel times. The bi-level optimization model in ref.~\cite{c32} considers passenger preferences such as departure times and ticket pricing strategies while optimizing the timetable to ensure both operator efficiency and passenger satisfaction.
\begin{equation}
\min \sum_{p \in \mathcal{P}} \left( c_p^{\text{wait}} + c_p^{\text{fare}} + c_p^{\text{con}} + c_p^{\text{delay}} \right)
\end{equation}
where $\mathcal{P}$ is the set of all passengers, $c_p^{\text{wait}}$ is the waiting cost, $c_p^{\text{fare}}$ is the ticket fare, $c_p^{\text{con}}$ is the congestion cost, and $c_p^{\text{delay}}$ is the penalty due to delay or transfer failure for passenger $p$.

Maximizing energy efficiency: with a growing emphasis on sustainability, timetable optimization often incorporates energy efficiency considerations. The energy savings achievable through timetable adjustment are inherently bounded by traffic constraints, and ESSs complement this by capturing residual RBE that scheduling alone cannot recover \cite{c16}.
\begin{equation}
\min \left( F + \theta \sum_{s \in \mathcal{S}} \pi_s F_s \right)
\end{equation}
where $F$ is the current-stage objective value, $F_s$ is the cost under scenario $s$, $\pi_s$ is the probability of scenario $s$, $\theta$ is the compensation weight, and $\mathcal{S}$ is the set of disruption scenarios.

Enhancing schedule robustness and service reliability: given the uncertainties in railway operations, robust optimization models aim to minimize schedule deviations caused by unpredictable disruptions. The data-driven distributionally robust optimization model \cite{c30} introduces stochastic and distributionally robust methods to account for uncertain travel times and delay propagation.
\begin{equation}
\min_{x \in \mathcal{X}} \; \mathcal{C}^{\text{plan}}(x) + \sup_{\mathbb{F} \in \mathcal{D}_\varepsilon(\widehat{\mathbb{F}})} \mathbb{E}_{\mathbb{F}} \left[ \mathcal{C}^{\text{rec}}(x, \xi) \right]
\end{equation}
where $x$ is the planned timetable, $\mathcal{X}$ denotes its feasible set satisfying operational and safety constraints, $\mathcal{C}^{\text{plan}}(x)$ represents the planned cost, $\xi$ denotes uncertain operational parameters, and $\mathcal{C}^{\text{rec}}(x, \xi)$ is the recovery cost. The ambiguity set $\mathcal{D}_\varepsilon(\widehat{\mathbb{F}})$ contains all distributions within Wasserstein distance $\varepsilon$ from the empirical distribution $\widehat{\mathbb{F}}$.

The following approaches have been proposed in recent studies regarding previous objectives. Firstly, synchronizing train acceleration and braking allows regenerative braking energy to be recovered and reused within the network, thereby enhancing overall energy utilization and reducing power consumption \cite{c33}. This synchronization requires coordinated timetable design across multiple trains, and its effectiveness depends on the availability of a receiving train at the right moment, a condition that cannot always be satisfied under real operational conditions. Secondly, minimizing peak power demand through multi-timescale operation strategies helps reduce stress on traction substations and grid infrastructure, contributing to more stable power delivery \cite{c31}. Thirdly, integrating train trajectory planning with timetable optimization enables smoother operations, reducing both mechanical wear and electrical energy consumption while maintaining punctuality \cite{c8}.

\subsubsection{Constraints}

To ensure the feasibility of train scheduling, various constraints must be considered. These constraints help maintain operational efficiency, safety, and reliability while optimizing resource utilization.

Departure and arrival time constraints: the timetable must ensure minimum travel times between stations while adhering to infrastructure and operational limits. The data-driven distributionally robust optimization model \cite{c30} accounts for these constraints by incorporating planned and actual timetables, ensuring adherence to predefined schedules while allowing for operational flexibility.
\begin{equation}
a_{i,j} - d_{i,j-1} \geq \underline{r}_{i,j} + \delta_{i,j}, \quad \forall (i,j)
\end{equation}
where $a_{i,j}$ and $d_{i,j-1}$ denote the actual arrival and departure times of train $i$ at station $j$ and its previous station, respectively; $\underline{r}_{i,j}$ is the minimum running time, and $\delta_{i,j}$ is the buffer time added for robustness.

Headway constraints: safety regulations require a minimum time interval between consecutive trains running on the same track to prevent conflicts and ensure safe braking distances. The multistage decision optimization approach \cite{c29} enforces minimum headway constraints.
\begin{equation}
d_{j,k} - d_{i,k} \geq H, \quad \forall (i,j) \in \mathcal{T}, k \in \mathcal{S}, i \neq j
\end{equation}
where $d_{i,k}$ is the departure time of train $i$ from segment $k$, and $H$ is the minimum headway required to ensure safe separation between trains.

Dwell time constraints: the duration of stops at stations must be managed to allow for efficient passenger boarding and alighting while preventing excessive dwell times that could disrupt schedule adherence. The dynamic metro train scheduling model \cite{add1} applies deep reinforcement learning (DRL) to reschedule train timetables in real time under disturbances, dynamically adjusting dwell and running times to minimize energy consumption.
\begin{equation}
\underline{\tau} \leq d_{i,j} - a_{i,j} \leq \bar{\tau}, \quad \forall i,j
\end{equation}
where $a_{i,j}$ and $d_{i,j}$ are the arrival and departure times of train $i$ at station $j$; $\underline{\tau}$ and $\bar{\tau}$ are the minimum and maximum allowable dwell times.

Track capacity and section occupancy constraints: trains cannot exceed the capacity limits of railway sections. Track occupancy constraints ensure that multiple trains do not simultaneously occupy the same track segment beyond its designed capacity. The multistage decision optimization model in ref.~\cite{c33} includes constraints to regulate track section occupancy and prevent conflicts in high-density railway corridors.
\begin{equation}
\sum_{i \in \mathcal{T}_s} \chi_{[a_{i,j}, d_{i,j}]}(t) \leq C_s, \quad \forall t, s
\end{equation}
where $\chi_{[a_{i,j}, d_{i,j}]}(t)$ is an indicator function that equals 1 if train $i$ occupies the section during time $t \in [a_{i,j}, d_{i,j}]$, and 0 otherwise.

Robustness constraints against disruptions: given the unpredictability of railway operations, robust optimization models incorporate buffer times to account for minor delays and prevent cascading disruptions. The data-driven distributionally robust timetabling model in \cite{c30} introduces uncertainty modeling to enhance resilience scheduling, ensuring that planned timetables remain stable even under fluctuating operational conditions.
\begin{equation}
a_{i,j} \geq \bar{a}_{i,j} + \theta_{i,j}, \quad \forall i,j
\end{equation}
where $\bar{a}_{i,j}$ is the scheduled arrival time at station $j$, and $\theta_{i,j}$ is a robustness buffer time to absorb minor delays and prevent propagation.

Dynamic scheduling problems in railway systems are addressed through various modeling approaches. Mathematical programming methods, including integer linear programming (IP)~\cite{c34}, mixed-integer linear programming (MILP)~\cite{c35}, and mixed-integer nonlinear programming (MINLP)~\cite{c36}, provide rigorous formulations for timetable optimization but are computationally intensive for large-scale instances. Simulation-based approaches~\cite{c37,c38} offer flexibility in modeling complex operational scenarios. Heuristic artificial intelligence methods such as branch-and-bound (B\&B) combined with beam search~\cite{c39} and neural network-based optimization~\cite{c40} improve scalability at the cost of optimality guarantees. Reinforcement learning (RL) has emerged as a powerful alternative for real-time scheduling. Multi-stage DRL~\cite{c41} enables online rescheduling for high-speed networks, while graph neural network (GNN)-based DRL~\cite{c42} extends this to large-scale disruption recovery. More recently, these approaches have been extended to jointly address scheduling and trajectory optimization within unified frameworks. In~\cite{c43}, a model is proposed using dueling double DQN (D3QN) combined with deep Q-learning from demonstrations (DQfD) to simultaneously generate rescheduled timetables and optimized train trajectories under disturbances, with computation time scaling linearly with problem size. A unified scheduling model is further developed \cite{c44} with long short-term memory (LSTM) networks and an attention mechanism, addressing both timetable optimization and rescheduling within a single policy network that generalizes to large-scale instances without retraining. Table~\ref{tab:scheduling_review} summarizes the literature on railway dynamic scheduling algorithms and operation control methods.

\setlength{\tabcolsep}{3pt}
\begin{longtable}{|>{\small\sffamily}p{1.464cm}|>{\small\sffamily}p{1.971cm}|>{\small\sffamily}p{2.253cm}|>{\small\sffamily}p{3.267cm}|>{\small\sffamily}p{1.690cm}|>{\small\sffamily}p{0.653cm}|>{\small\sffamily}p{1.070cm}|>{\small\sffamily}p{0.811cm}|>{\small\sffamily}p{1.295cm}|}
\caption{Scheduling Algorithms and Operation Control Methods}
\label{tab:scheduling_review}\\
\hline
\textbf{Category} & \textbf{Method} & \textbf{Algorithm} & \textbf{Main Contribution} & \textbf{Train Type} & \textbf{Ref.} & \textbf{Comp.} & \textbf{RT} & \textbf{Robust.} \\
\hline
\endfirsthead
\multicolumn{9}{l}{\small\itshape (Table~\ref{tab:scheduling_review} continued)}\\
\hline
\textbf{Category} & \textbf{Method} & \textbf{Algorithm} & \textbf{Main Contribution} & \textbf{Train Type} & \textbf{Ref.} & \textbf{Comp.} & \textbf{RT} & \textbf{Robust.} \\
\hline
\endhead
\hline
\multicolumn{9}{r|}{\small\itshape Continued on next page}\\
\endfoot
\hline
\multicolumn{9}{|l|}{\footnotesize Comp. = Computational complexity; RT = Real-time capability; Robust. = Robustness to uncertainty} \\
\hline
\endlastfoot
Scheduling algorithms
 & \multirow{3}{=}{\raggedright Mathematical programming}
 & Integer linear programming & Design train-level Pareto solution for energy-efficient timetable & High-speed & \cite{c34} & High & No & Low \\ \cline{3-9}
 & & MILP & Maximize resilience against disruptions & Urban rail & \cite{c35} & High & No & Medium \\ \cline{3-9}
 & & MINLP + rolling horizon & Enhance computational efficiency for passenger-freight collinear railway & Mixed & \cite{c36} & High & No & Low \\ \cline{2-9}
 & \multirow{2}{=}{\raggedright Simulation}
 & Event-based simulation & Improve train management system & Urban rail & \cite{c37} & Medium & Partial & Low \\ \cline{3-9}
 & & Discrete-event simulation & Generate operation plan under disruptions considering headway regularity index & Urban rail & \cite{c38} & Medium & Partial & Low \\ \cline{2-9}
 & \multirow{2}{=}{\raggedright AI (Heuristic)}
 & B\&B + beam search & Design large-scale robust timetabling algorithm & Conventional & \cite{c39} & High & No & Low \\ \cline{3-9}
 & & Neural network heuristic & Integrate neural networks for RBE optimization with variable headway & Urban rail & \cite{c40} & Medium & Partial & Low \\ \cline{2-9}
 & \multirow{4}{=}{\raggedright AI (RL)}
 & Multi-stage DRL & RL-based real-time rescheduling for high-speed rail networks & High-speed & \cite{c41} & Low & Yes & Medium \\ \cline{3-9}
 & & GNN + DRL & Restore train operation after large-scale disruptions & Conventional & \cite{c42} & Low & Yes & Medium \\ \cline{3-9}
 & & D3QN + DQfD & Simultaneously generate rescheduled timetable and optimized trajectory; linear computation scaling & High-speed & \cite{c43} & Medium & Partial & Medium \\ \cline{3-9}
 & & DRL (LSTM + attention) & Unified timetable optimization and rescheduling; scalable without retraining & High-speed & \cite{c44} & Low & Yes & Medium \\ \hline
Operation control
 & \multirow{4}{=}{\raggedright Model Predictive Control (MPC)}
 & MPC & Online energy-saving speed profile generation & High-speed & \cite{c6} & Medium & Yes & Low \\ \cline{3-9}
 & & Stability-enhanced MPC & Safe following distances and smooth acceleration profiles & Urban rail & \cite{c45} & Medium & Yes & Low \\ \cline{3-9}
 & & Nonlinear safety MPC & Reduce inter-train spacing on gradient terrain safely & Conventional & \cite{c46} & Medium & Yes & Medium \\ \cline{3-9}
 & & Cooperative MPC & Multi-train coordination for high network capacity & High-speed & \cite{c47} & High & Partial & Low \\ \cline{2-9}
 & \multirow{2}{=}{\raggedright Fuzzy Logic Control}
 & FPC & Optimize traction and braking via adaptive fuzzy rules & High-speed & \cite{c48} & Low & Yes & Medium \\ \cline{3-9}
 & & T-S Fuzzy Bilinear Model-Based Control & Online trajectory regulation for high-speed trains & High-speed & \cite{c49} & Low & Yes & Medium \\ \cline{2-9}
 & \multirow{4}{=}{\raggedright AI (RL)}
 & DDPG & Balance punctuality and energy savings in trajectories & High-speed & \cite{c50} & Low & Yes & Medium \\ \cline{3-9}
 & & DQN & Enhance ATO scheduling accuracy and energy efficiency & Urban rail & \cite{c51} & Low & Yes & Low \\ \cline{3-9}
 & & TD3 & Real-time energy-efficient trajectory generation under disturbances & Urban rail & \cite{c52} & Low & Yes & Medium \\ \cline{3-9}
 & & TPPA & Real-time speed profile re-optimization; reduces complexity from O($n^2$) to O($n$) & Conventional & \cite{c53} & Low & Yes & Low \\
\end{longtable}
\setlength{\tabcolsep}{6pt}

\subsection{Trajectory Optimization in Railway Networks}

Operation control of high-speed trains focuses on optimizing train speed trajectories to ensure safe, reliable, and energy-efficient operations \cite{c54,c55}. The research in this field encompasses both centralized and distributed control strategies, leveraging advanced optimization techniques such as Model Predictive Control, fuzzy control, and Reinforcement Learning (RL) methods.

MPC has become a cornerstone of railway operation control, offering a structured framework for balancing multi-objective optimization while addressing real-time constraints. Stability-enhanced MPC ensures safe following distances and smooth acceleration profiles, improving safety in urban rail transit \cite{c45}, while Nonlinear Safety Equilibrium Spacing-Based MPC enhances virtually coupled train operations over gradient terrains, reducing inter-train spacing without compromising safety \cite{c46}. Additionally, energy-efficient MPC dynamically adjusts speed trajectories to minimize energy consumption in high-speed rail networks \cite{c6}. The resulting speed profiles directly determine the magnitude and distribution of regenerative braking energy, which in turn shapes the power and capacity requirements of onboard or wayside ESSs.

To further optimize network-wide efficiency, cooperative control strategies are increasingly employed. Cooperative MPC enables synchronized multi-train coordination, minimizing energy use while maintaining optimal headway adherence, which is particularly beneficial in dense rail traffic scenarios \cite{c47}. 
Fuzzy logic approaches effectively handle operational uncertainties. Fuzzy Predictive Control (FPC) enhances locomotive energy efficiency by optimizing traction and braking based on adaptive fuzzy rules \cite{c48}, while T-S Fuzzy Bilinear Model-Based Control integrates predictive logic for real-time trajectory regulation in high-speed rail \cite{c49}.

Incorporating artificial intelligence has further revolutionized train control by enabling real-time adaptability and enhanced efficiency. RL-based approaches, such as Deep Deterministic Policy Gradient (DDPG), optimize high-speed train trajectories to balance punctuality and energy savings \cite{c50}, while DQN enhances ATO by improving scheduling accuracy and energy efficiency in urban rail \cite{c51}. Moreover, twin-delayed deep deterministic policy gradient (TD3) handles uncertain disturbances during train operation and generates optimal energy-efficient trajectories online \cite{c52}. Beyond these learning-based methods, a tour-adaptive partial-bounding pulse algorithm (TPPA) is proposed in \cite{c53} to reformulate energy-efficient speed profile optimization as a shortest path problem with time windows, reducing computational complexity from $\mathcal{O}(n^2)$ to $\mathcal{O}(n)$ and enabling real-time re-optimization onboard under both plan deviations and speed limit changes.

A comparison of the three major control methods reveals distinct trade-offs across computational complexity, real-time capability, and robustness to uncertainty. MPC offers a principled framework for multi-objective optimization with explicit constraint handling, and is well-suited to deterministic or mildly uncertain operating conditions. However, its computational burden grows with the prediction horizon and the number of trains, limiting scalability in large networks \cite{c6}. Fuzzy logic control, by contrast, requires no precise system model and delivers low computational cost with strong real-time capability, making it practical for onboard deployment, though its performance depends heavily on rule design and it struggles to guarantee global optimality \cite{c48,c49}. RL methods offer the greatest adaptability to complex and stochastic environments, learning control policies directly from interaction without requiring explicit dynamic models; yet their training data requirements are substantial, interpretability remains limited, and deployment in safety-critical railway systems demands careful validation \cite{c50,c51,c52}. While RL has demonstrated strong results in urban rail contexts, its application to high-speed mainline railways with more complex energy and safety constraints remains largely unexplored. Across all three methods, ESS state of charge (SOC) is rarely incorporated as part of the control input, and the interaction between speed trajectory decisions and ESS operation remains largely unaddressed.

By integrating predictive control, cooperative optimization, and AI-driven adaptive strategies, modern railway systems can achieve enhanced operational efficiency, improved passenger experience, and greater sustainability. 

%%=============================================
\section{Energy Storage System Management and Control for Railway Systems}
\label{sec:ess}
%%=============================================

ESSs are essential for improving both energy utilization and operational reliability in electrified railway networks. In sloped sections, downhill trains produce surplus energy through regenerative braking, which is typically fed back into the overhead catenary. However, due to limited grid absorption capacity and unsynchronized train movements, much of this energy is underutilized or rejected \cite{c56}. 
ESSs help address this issue by locally capturing excess energy for later use. The extent of this mismatch is directly shaped by timetable decisions and speed profiles, and ESS sizing and control strategies therefore need to reflect the operational patterns determined at the scheduling and trajectory level. Recent studies further indicate that degradation-aware operation, capacity configuration, and hybrid storage coordination are becoming important design considerations for railway ESS deployment~\cite{c_add12}.

\begin{figure*}[htbp]
\centering
\includegraphics[width=0.82\textwidth]{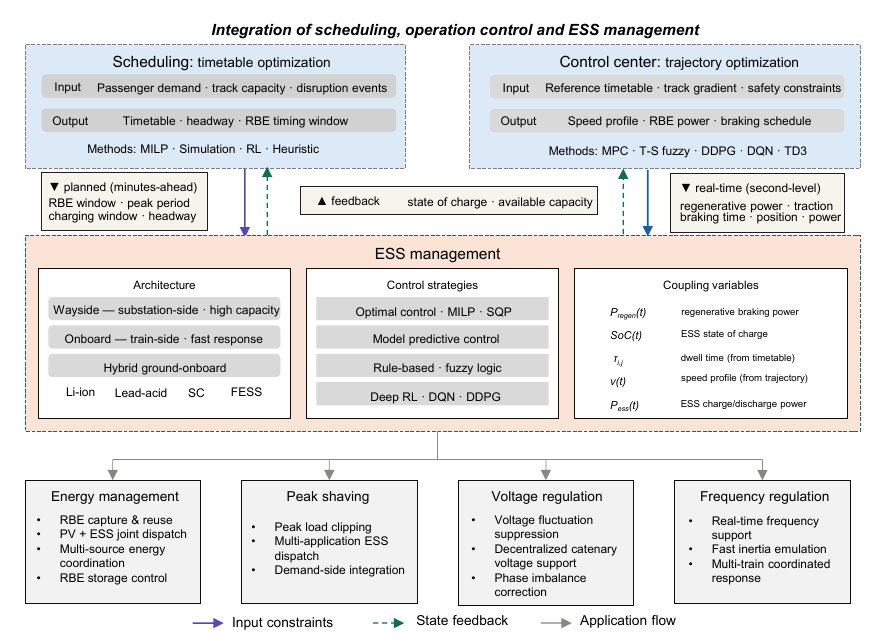}
\caption{Integration framework of scheduling, operation control, and ESS management in electrified railway systems.}
\label{fig:ess_framework}
\end{figure*}

As illustrated in Fig.~\ref{fig:ess_framework}, the ESS management layer receives two categories of information from the upper layers. From the scheduling center, planned information including RBE timing windows, charging opportunity windows, and peak demand periods is provided on a minutes-ahead basis to support day-ahead energy planning. From the control center, real-time signals including regenerative braking power, traction demand, and braking schedules are transmitted at the second level to enable real-time charge and discharge control. In return, ESS SOC and available capacity are fed back to both upper layers, allowing the scheduler to adjust the next timetable cycle and enabling the trajectory controller to modify speed profiles when storage capacity is constrained.

\begin{figure}[tbp]
\centering
\includegraphics[width=0.55\textwidth]{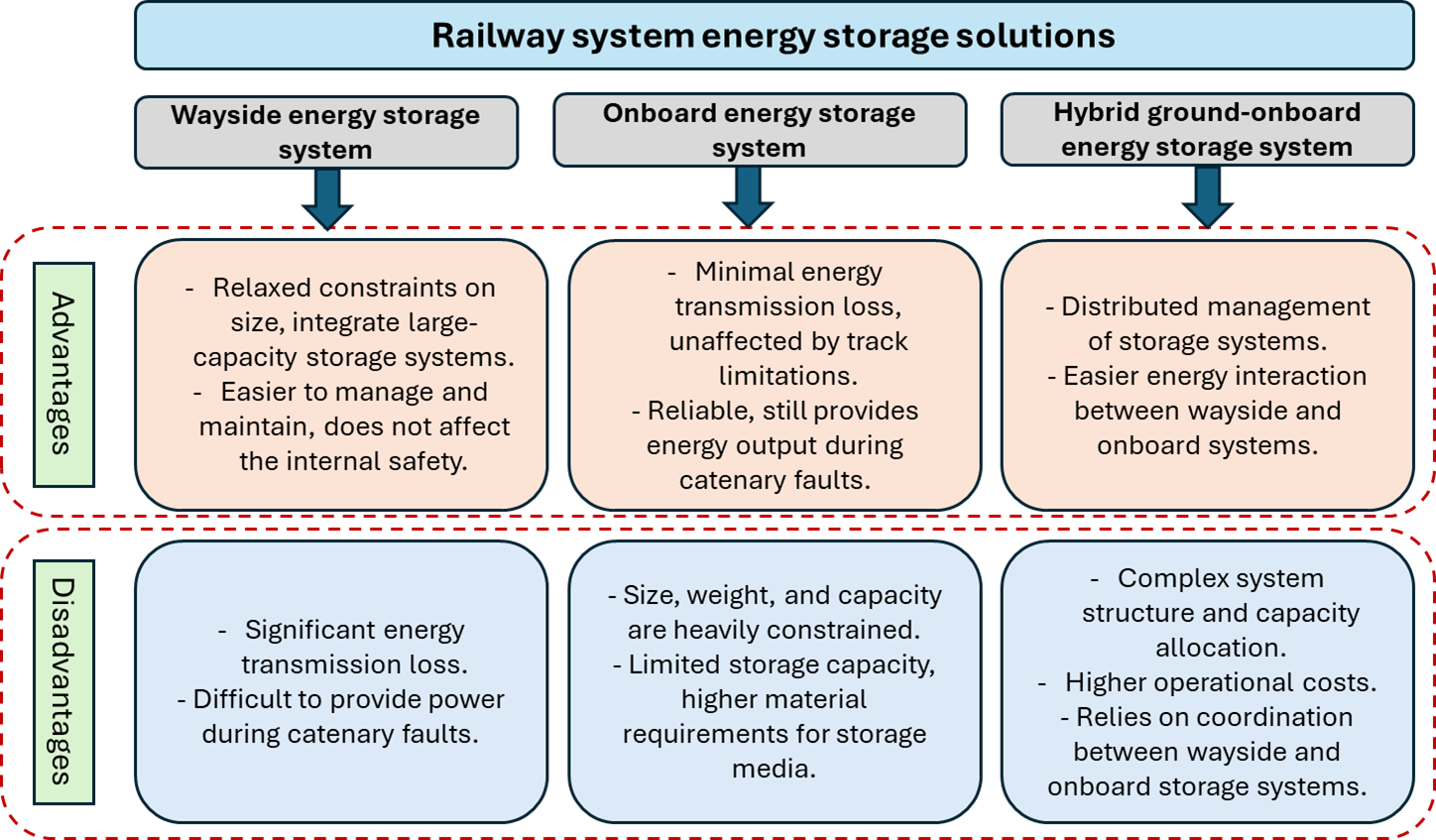}
\caption{Advantages and disadvantages of railway system integrated ESSs.}
\label{fig:ess_solution}
\end{figure}

ESSs in railway systems can be categorized into three main types based on the installation location of the storage devices: wayside ESSs \cite{c57}, onboard ESSs \cite{c58}, and hybrid ground-onboard ESSs. These topologies differ significantly in hardware structures, capacity allocation methods, and energy management strategies, making it essential to analyze their unique characteristics and applications. Fig.~\ref{fig:ess_solution} shows the summarized solutions of advantages and disadvantages of railway system integrated ESSs.

1) Wayside ESSs store braking energy by installing storage devices at substations, offering a centralized solution to manage energy flow. These systems typically connect to the traction power supply network and are designed to capture energy generated by regenerative braking. ~\cite{c24} focused on the optimal control of reversible substations and wayside storage devices for energy savings and voltage stabilization, proposing an optimization strategy based on the linearization of DC power flow equations solved via a successive approximation method. More recent studies have extended wayside ESS research toward flexible hybrid energy storage system (HESS) 
operation~\cite{c_add13}, railway power conditioner (RPC)-based RBE 
utilization~\cite{c_add14, c59}, and techno-economic capacity design for AC traction power supply systems~\cite{c_add15}. The effectiveness of such control strategies depends on the predictability of train power demand patterns, which are governed by the timetable and speed profiles adopted at the operational level.
Additionally, some designs integrate regenerative energy feedback systems by connecting an inverter \cite{c56}, which supplies power to station facilities and signaling equipment. While this approach extends functionality, frequent load fluctuations in railway systems may impact the safety and stability of the connected power grid.

2) Onboard ESSs integrate energy storage devices directly onboard, providing a decentralized and highly responsive energy management solution. These systems are designed to handle RBE locally, eliminating the transmission losses associated with traction grid systems. For instance, the optimal sizing of onboard energy storage devices to minimize catenary energy consumption is addressed in \cite{c60}, while \cite{c61} extends this to hybrid onboard configurations considering long-term train operation. 
In addition to sizing-oriented studies, adaptive eco-driving has been investigated for electric trains with onboard energy storage devices, showing that storage operation can be jointly considered with driving strategy and feasibility constraints~\cite{c_add16}.
Beyond energy recovery, onboard ESSs also provide a function that enhances system resilience. In the event of regional faults in the traction power network, trains equipped with onboard storage can continue operating under emergency power, whereas those without backup systems may become stranded and pose operational and safety risks \cite{c58}. However, onboard systems face significant challenges due to the limited space and weight constraints of train cars, which restrict the capacity and scalability of storage devices.

\begin{figure}[tbp]
\centering
\includegraphics[width=0.45\textwidth]{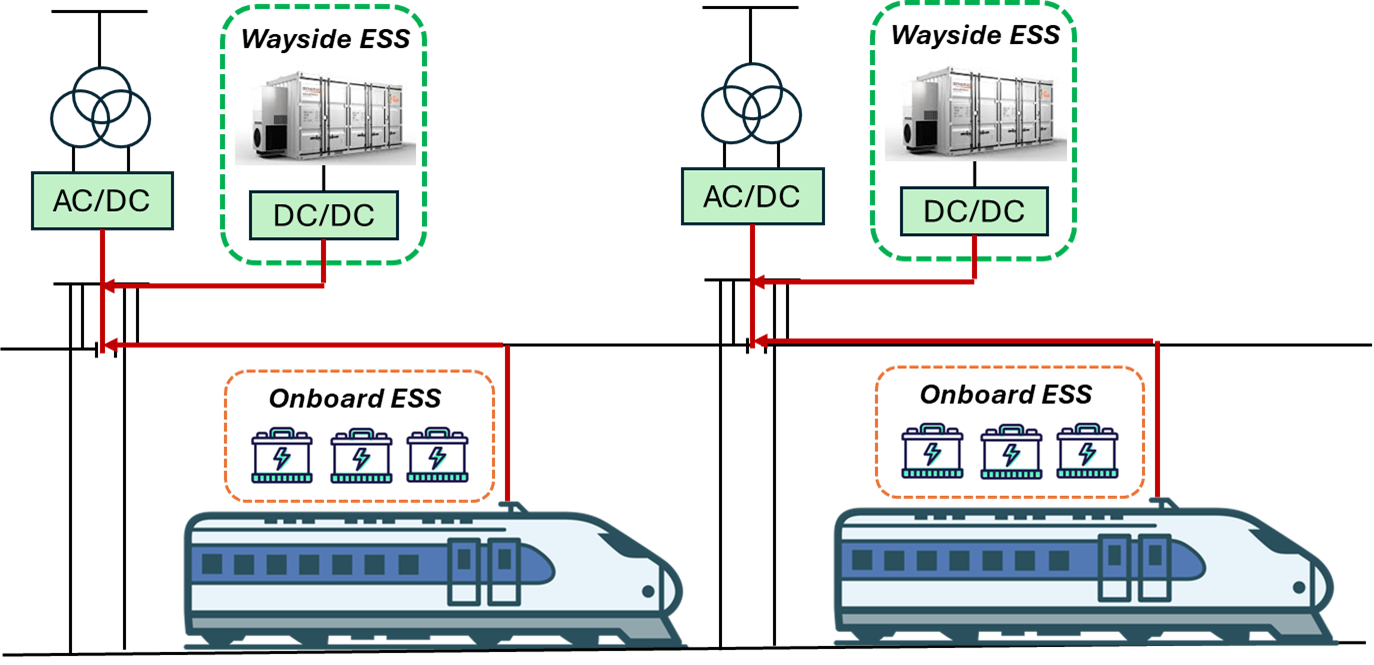}
\caption{Diagram of hybrid ESS configuration.}
\label{fig:hybrid_ess}
\end{figure}

3) Hybrid ground-onboard ESSs combine the strengths of wayside and onboard systems, making them a critical component of integrated energy solutions for railways. In this configuration, the high-power output of onboard storage devices complements the high-capacity storage capabilities of wayside systems, creating a cooperative approach to energy management. 
Bilevel sizing and control strategies have also been proposed for battery--ultracapacitor HESSs in urban rail transit, where both operating cost and substation stability are considered~\cite{c_add12}.
Fig.~\ref{fig:hybrid_ess} shows the hybrid ground-onboard ESS configuration \cite{c62,a0}. This hybrid configuration reduces the capacity requirements for onboard systems, minimizes energy transmission losses, and allows for independent operation of each storage system. Wayside systems focus on long-term energy management and grid stability, while onboard systems provide localized high-power support. In fault conditions, the hybrid setup can dynamically adjust the relationship between wayside and onboard storage systems to enhance emergency power supply capabilities.

%%=============================================
\section{Application Scenarios and Advanced Functionalities for Railway Systems}
\label{sec:application}
%%=============================================

This section examines how these systems support energy management, strengthen grid resilience, and improve overall operational performance. Table~\ref{tab:application} shows the application scenarios and advanced functionalities for railway system.

\setlength{\tabcolsep}{3pt}
\begin{longtable}{|>{\small\sffamily}p{1.614cm}|>{\small\sffamily}p{2.536cm}|>{\small\sffamily}p{4.611cm}|>{\small\sffamily}p{1.729cm}|>{\small\sffamily}p{1.498cm}|>{\small\sffamily}p{2.305cm}|>{\small\sffamily}p{0.634cm}|}
\caption{Application Scenarios and Advanced Functionalities}
\label{tab:application}\\
\hline
\textbf{Scenario} & \textbf{Method} & \textbf{Main Contribution} & \textbf{ESS Type} & \textbf{Train Type} & \textbf{ESS Coupling} & \textbf{Ref.} \\
\hline
\endfirsthead
\multicolumn{7}{l}{\small\itshape (Table~\ref{tab:application} continued)}\\
\hline
\textbf{Scenario} & \textbf{Method} & \textbf{Main Contribution} & \textbf{ESS Type} & \textbf{Train Type} & \textbf{ESS Coupling} & \textbf{Ref.} \\
\hline
\endhead
\hline
\multicolumn{7}{r|}{\small\itshape Continued on next page}\\
\endfoot
\hline
\multicolumn{7}{|l|}{\footnotesize ESS Coupling: Coupling level (Joint / Coordinated / Decoupled) / ESS variable (SOC, Power, None) / Timescale (Real-time, Day-ahead)} \\
\hline
\endlastfoot
Energy management
 & MILP & Balance energy supply and demand considering PV generation and storage uncertainties & Wayside & Urban rail & Coordinated / Power / Day-ahead & \cite{c63} \\ \cline{2-7}
 & Two-level optimization & Centralized day-ahead and decentralized real-time coordination of trains, wayside storage, and distributed sources & Wayside & Conventional & Coordinated / Power / Day-ahead & \cite{c64} \\ \cline{2-7}
 & Sequential quadratic programming & Hierarchical EMS for networked flexible traction substations to enhance RBE and PV utilization & Wayside & Conventional & Coordinated / Power / Day-ahead & \cite{c65} \\ \cline{2-7}
 & DRL (double DQN) & Multi-timescale reward-based strategy maximizing RBE utilization modeled as Markov decision process & Wayside & Urban rail & Joint / SOC / Real-time & \cite{c66} \\ \cline{2-7}
 & Convex optimization & Two-step joint optimization of train operation, timetable, and onboard ESD management to minimize net energy consumption & Onboard & Urban rail & Joint / SOC+Power / Day-ahead & \cite{c58} \\ \cline{2-7}
 & MILP + Monte Carlo & Concurrent optimization of train trajectory and onboard ESS under stochastic RBE distribution & Onboard & Urban rail & Joint / SOC+Power / Real-time & \cite{c16} \\ \cline{2-7}
 & Nonlinear programming & Unified model integrating traction dynamics, battery behavior, and RBE recovery; validated on Jerusalem Light Rail & Onboard & Light rail & Joint / SOC+Power / Real-time & \cite{c73} \\ \cline{2-7}
 & Fuzzy logic control & Adaptive battery-SC power sharing considering headway-dependent RBE to delay battery degradation & Hybrid & Urban rail & Joint / SOC+Power / Real-time & \cite{c74} \\ \hline
Peak shaving and load balancing
 & DRL & Optimize energy utilization addressing thermal constraints of power electronics and battery modules & Wayside & Conventional & Joint / SOC+Power / Real-time & \cite{c67} \\ \cline{2-7}
 & Particle Swarm Optimization (PSO)-based optimization & Multi-application strategy simultaneously achieving peak clipping and RBE utilization with ESS-RPC & Wayside & High-speed & Joint / SOC+Power / Day-ahead & \cite{c69} \\ \cline{2-7}
 & Three-stage MPC & Day-ahead MILP scheduling, intra-day rolling MPC correction, and real-time power quality control for FTSS & Wayside & Conventional & Joint / SOC+Power / Day-ahead & \cite{c77} \\ \hline
Voltage and frequency regulation
 & MILP & ESS absorbs RBE and releases it to prevent voltage sag under dynamic unbalanced load conditions & Wayside & High-speed & Coordinated / Power / Real-time & \cite{c70} \\ \cline{2-7}
 & MPC & Decentralized voltage regulation and inertia support among multiple traction substations in MVdc systems & Wayside & High-speed & Coordinated / Power / Real-time & \cite{c71} \\ \cline{2-7}
 & Successive approximation & Optimal control of reversible substations and wayside storage for voltage stabilization and energy savings & Wayside & Urban rail & Coordinated / Power / Real-time & \cite{c24} \\ \cline{2-7}
 & Sequential cutting-plane & Day-ahead capacity estimation with real-time coordination for primary frequency regulation from trains & / & High-speed & Decoupled / None / Real-time & \cite{c25} \\ \cline{2-7}
 & Successive quadratic programming & Adaptive frequency control by dynamically adjusting train power consumption via online trajectory optimization & / & High-speed & Decoupled / None / Real-time & \cite{c26} \\
\end{longtable}
\setlength{\tabcolsep}{6pt}

\subsection{Energy Management in Railway Systems}

Energy management in railway systems targets the effective utilization of RBE, integration of renewable energy sources such as PV generation, and optimization of power loads to improve overall system efficiency. The RBE available for capture is governed by train speed profiles and timetable design, and effective energy management strategies therefore require train operational data as a critical input. A stochastic energy management model is presented in \cite{c63} for smart railway stations that integrates regenerative braking, ESSs, and PV generation, effectively balancing supply and demand under uncertainty while reducing energy losses and costs. Recent station-level studies have further extended this framework to charging coordination under peak load constraints, where RBE, renewable generation, ESS operation, and charging flexibility are jointly scheduled~\cite{c_add18}. Beyond conventional station loads, RBE has also been explored for integration with EV fast charging infrastructure, enabling synergistic dispatch between railway braking energy and charging demand~\cite{c68}.
To extend beyond single station control, a two-level optimization framework combining centralized day-ahead scheduling with decentralized real-time control is introduced in \cite{c64}, coordinating trains, wayside storage, and distributed sources for energy savings and load balancing. Though validated on the Malaga-Fuengirola line, its scalability to multi-line systems remains a challenge.

To scale this coordination across spatially distributed substations, a hierarchical energy management strategy for networked flexible traction substations is proposed in \cite{c65} to enhance RBE and PV energy utilization in AC railway systems. 
Recent robust and real-time formulations have further considered traction load and PV uncertainties, energy storage aging, and railway grid service provision in flexible traction substation operation~\cite{c_add17,c_add20}.
By employing unified power flow controllers at sectioning posts, the system enables power exchange between adjacent substations, reducing capacity requirements and enhancing the utilization of RBE and PV energy. The two-layer control framework consists of a global level for optimizing energy exchange and peak shaving, and a local level where each substation autonomously manages power output, storage, and power quality. This approach improves system flexibility and reliability, though real-time coordination and scalability remain an implementation challenge. The joint optimization of wayside ESS sizing, siting, and operation together with train trajectory is investigated in \cite{a2}, where a coordinated planning model minimizes total electricity cost and is validated on the Swedish railway network. Meanwhile, machine learning methods offer promising alternatives. A DRL-based energy management strategy for RBE storage systems in railway power systems is proposed in~\cite{c66}. The multiobjective problem of maximizing regenerative energy utilization and reducing power demand is modeled as a Markov decision process and solved using double DQN, with a multistage reward function coordinating learning across different timescales. Building on 
this, SOC-aware RBE management has recently been emphasized to avoid storage saturation and improve the coordination between regenerative braking absorption and battery operating limits~\cite{c_add19}.

Beyond wayside-focused strategies, onboard and hybrid ESS configurations introduce additional degrees of freedom in energy management. As discussed in Section~\ref{sec:ess}, joint optimization of train operation and onboard management has been demonstrated to significantly reduce net energy consumption \cite{c58}, with further extensions to stochastic RBE environments \cite{c16}. An integrated optimization framework for light rail systems is proposed in \cite{c73}, incorporating detailed traction dynamics, battery behavior, and infrastructure constraints into a unified nonlinear model. Validation on the Jerusalem Light Rail Green Line demonstrates that cost minimization with combined regenerative braking and ESS yields a more balanced trade-off between performance and energy expenditure than time minimization alone. For hybrid ESS configurations combining battery and supercapacitor at the wayside, an adaptive fuzzy logic control strategy is proposed in \cite{c74} to distribute power between the two storage media while accounting for the influence of headway on RBE availability, effectively delaying battery degradation through reduced charge and discharge stress.

Comparing the ESS coupling characteristics across these methods reveals an important structural divide that is increasingly shaped by uncertainty modeling, SOC-aware storage operation, and station-level demand flexibility. Wayside-focused strategies such as \cite{c63}, \cite{c64}, \cite{c65} operate in a coordinated mode, where ESS responds to upper-level scheduling signals but is not a joint decision variable. This decoupling simplifies implementation but limits the achievable energy savings. In contrast, onboard-oriented methods \cite{c16}, \cite{c58}, \cite{c73} and hybrid approaches \cite{c74,a1} achieve joint optimization by incorporating SOC directly into the objective function, enabling tighter energy recovery but at the cost of greater computational and modeling complexity. The DRL approach in \cite{c66} occupies an intermediate position, treating SOC implicitly through the reward structure rather than as an explicit constraint. Across all reviewed methods, scalability to multi-line high-speed networks and real-time deployment under communication constraints remain open challenges.

\subsection{Peak Shaving and Load Balancing in Railway Systems}

Peak shaving and load balancing are critical for improving the operational efficiency and stability of railway power supply systems. With increasing energy demands and load variations, strategies have been developed to address peak power consumption, optimize energy flow, and alleviate stress on traction power networks. 
Early approaches focused on transferring traction energy across time using battery ESS to reduce the peak power drawn from traction substations~\cite{c_add22, c_add24}. 
Building on this, coordinated demand response strategies have been proposed to jointly adjust train load, ESS output, and passenger comfort constraints, enabling more flexible peak management across the network~\cite{c_add23}. To handle additional operational 
constraints, a thermal-constrained optimization approach is proposed in~\cite{c67}, incorporating ESSs and a DQN to manage power flow while accounting for thermal limits in power electronics and battery modules. A PSO-based multi-application strategy in~\cite{c69} simultaneously achieves peak load clipping and RBE utilization to improve the comprehensive economic benefits of the traction system. To address the uncertainty of both PV generation and traction load in flexible traction substations, a three-stage optimization strategy is proposed in~\cite{c77}, in which a day-ahead MILP model plans 
ESS charging and discharging to minimize daily operational costs, an intra-day adaptive MPC stage corrects the plan against forecast errors, and a real-time stage suppresses negative sequence currents using the remaining power conditioner capacity. This 
multi-timescale structure achieves joint optimization of SoC trajectories and power quality, demonstrating stronger performance than conventional single-stage approaches.

Comparing the ESS coupling depth across these methods reveals a clear progression. The earlier strategies~\cite{c_add22, c_add24} treat ESS primarily as a passive buffer, operating in a coordinated mode without explicit SOC optimization. The demand response approach in~\cite{c_add23} introduces tighter coupling by linking ESS dispatch 
to train operation decisions, though scheduling-level variables remain decoupled. The DQN approach in~\cite{c67} and the three-stage MPC in~\cite{c77} both incorporate SOC as a joint decision variable in real-time control, while the PSO-based strategy 
in~\cite{c69} optimizes ESS capacity and dispatch on a day-ahead basis with SOC constraints explicitly enforced. Nevertheless, existing methods predominantly address wayside ESS in isolation from train scheduling. The substitution relationship between 
ESS peak shaving capacity and timetable-level energy synchronization warrants closer attention, since adequate ESS capacity can relax the energy coordination constraints placed on train scheduling, while well-designed timetables can in turn reduce ESS capacity requirements.

\subsection{Voltage and Frequency Regulation in Railway Systems}

Ensuring stable voltage and frequency regulation is essential for the reliable operation of electrified railway systems, especially as power converter integration expands. Voltage regulation maintains power quality by keeping supply within acceptable limits, mitigating issues like voltage drops, fluctuations, and imbalances caused by variable train loads and renewable inputs. Frequency regulation, meanwhile, balances power generation and demand at the grid level. Electrified railways can actively support this by adjusting power consumption or leveraging ESSs for fast response. With the growing share of renewables and their inherent volatility, such participation becomes increasingly valuable. By contributing to primary frequency control and fast grid response, railway systems enhance both operational reliability and broader grid stability.

1)\quad\textit{Voltage Regulation}: Recent studies have focused on advanced voltage regulation strategies using ESSs and power electronics to address catenary voltage instability and power quality issues. A real-time strategy that integrates ESS with power flow control to mitigate dynamic voltage unbalance is proposed in \cite{c70}. A predictive model adjusts power flow based on train load, while the ESS absorbs regenerative energy and releases it to prevent voltage drops. Power flow controllers further optimize energy dispatch, reducing system losses and improving resilience. A virtual inertia control strategy is designed in \cite{c71} to provide inertia support and realize decentralized voltage regulation among different TPSSs. Ref.~\cite{c24} addresses voltage stabilization through optimal control of reversible substations and wayside storage. This approach enhances energy recovery and RBE utilization while maintaining voltage stability, with validation on the Thessaloniki metro network. However, its DC-specific focus does not address challenges in AC systems, such as reactive power control and phase imbalance. For AC railways, a power electronic autotransformer solution is introduced in \cite{c27} using back-to-back voltage source converters to correct voltage imbalances from split-phase configurations. The system improves power factor, enables reactive power control, and balances grid currents. Simulations and real-time tests confirm its effectiveness, though high cost and system complexity pose barriers to large-scale deployment.

2)\quad\textit{Frequency Regulation}: With growing renewable energy integration, electrified railways are increasingly seen as active participants in grid frequency regulation through adaptive energy management. A framework enabling electric trains is proposed in \cite{c25} to participate in primary frequency control by combining day-ahead capacity estimation with real-time response coordination. A sequential cutting-plane algorithm effectively addresses nonlinear estimation challenges. While promising, the approach faces difficulties in coordinating multiple trains in dense networks. To enhance responsiveness, an adaptive strategy using online trajectory optimization and droop control is introduced in \cite{c26}. This allows trains to support grid frequency without compromising schedules. When ESS regulation is applied alongside trajectory adjustment, the two mechanisms address different timescales of frequency deviation, with ESS responding to rapid fluctuations and speed profile modification providing more sustained power adjustment. Though effective individually, coordinating both within a unified control framework remains an open challenge, particularly in complex railway environments with dense traffic. In parallel, grid-side solutions are introduced in \cite{c72} by integrating virtual inertia into railway power conditioners to improve frequency stability in renewable-rich railway systems.

The coordinated use of voltage and frequency regulation strategies can support grid stability, but the extent of this contribution depends on specific system designs and the level of interaction permitted by grid operators. From an ESS coupling perspective, the reviewed voltage regulation methods operate primarily in a coordinated mode, where ESS responds to real-time power flow signals without explicit SOC optimization in the scheduling layer. This limits the ability to pre-position ESS state ahead of anticipated demand patterns. Frequency regulation methods \cite{c25,c26} are decoupled from ESS entirely, relying instead on train load flexibility, which avoids storage investment but introduces constraints on timetable adherence. A unified framework that coordinates ESS dispatch, train speed profiles, and frequency response across multiple timescales remains an open research challenge, particularly as renewable penetration increases the volatility of grid frequency deviations. With proper integration of energy storage, adaptive control, and power electronics, railways can enhance power quality and provide limited grid services. Future research should focus on scalable, real-time control frameworks that account for system constraints and coordination with power networks.

%%=============================================
\section{Conclusion}
\label{sec:conclusion}
%%=============================================

This review has systematically analyzed the state of art developments in the optimization of electrified railway systems, with a focus on dynamic scheduling, operation control, and the integration of ESSs. The deployment of advanced optimization algorithms enables timetable adjustments, multi-train coordination, and energy-efficient trajectory planning under complex operational constraints. These methods contribute to minimizing total energy consumption, reducing delays, and enhancing the efficiency of railway operations. Furthermore, the integration of ESSs has proven essential for capturing RBE and enhancing power supply reliability. A cross-architecture comparison of wayside, onboard, and hybrid ESS configurations indicates that system-level performance is highly dependent on precise sizing, placement, and control coordination, particularly under dynamic load and voltage conditions. These architectures have been applied across energy management scenarios, including load leveling, peak shaving, and real-time voltage and frequency regulation, thereby supporting both operational efficiency and grid stability across diverse railway scenarios. Future research should focus on the holistic integration of train operation control, dynamic scheduling, and ESS management under unified, data-driven optimization frameworks to realize both energy-optimal and grid-responsive system.

\bibliographystyle{cas-model2-names}

\end{document}